\documentclass{elsart}
\usepackage{amssymb}
\usepackage{latexsym}
\usepackage{graphics}
\usepackage{graphicx}
\usepackage{bm}% bold math

\begin{document}
%%%%%%%%%%%%%%%%%%%%%%%%%%%%%%%%%%%%%%%%%%%%%%
\begin{frontmatter}

\title{The distribution of 
first-passage times and durations in FOREX and future markets}

\author[label1]{Naoya Sazuka}
\ead{Naoya.Sazuka@jp.sony.com}
\address[label1]{System Technologies Laboratories, 
Sony Corporation, 
5-1-12 Kitashinagawa Shinagawa-ku, 
Tokyo 141-0001, Japan} 

\author[label2]{Jun-ichi Inoue}
\ead{j$\underline{\,\,\,}$inoue@complex.eng.hokudai.ac.jp}
\address[label2]{Complex Systems Engineering, 
Graduate School of Information 
Science and Technology, 
Hokkaido University, 
N14-W9, Kita-ku, Sapporo 060-0814, Japan}

\author[label3]{Enrico Scalas}
\ead{enrico.scalas@mfn.unipmn.it}
\address[label3]{Dipartimento di Scienze e Tecnologie Avanzate, 
Universit\`a del Piemonte Orientale, 
Via Bellini, 25 g I-15100 
Alessandria, Italy}

%%%%%%%%%%%%%%%%%%%%%%%%%%%%%%%%%%%%%%%%%%%%%%%%%%%%%%%%%%%%%%%%%%%
%%                         Abstract                              %%
%%%%%%%%%%%%%%%%%%%%%%%%%%%%%%%%%%%%%%%%%%%%%%%%%%%%%%%%%%%%%%%%%%%
\begin{abstract}
Possible distributions are discussed
for intertrade durations and first-passage processes 
in financial markets. The view-point of {\em renewal theory}
is assumed. 
In order to represent market data with relatively long durations, two types of 
distributions are used, namely, a distribution derived 
from the so-called Mittag-Leffler survival function and 
the Weibull distribution. 
For Mittag-Leffler type distribution,
the average waiting time (residual life time) 
is strongly dependent on the choice of a cut-off 
parameter $t_{\rm max}$, whereas the results based on the Weibull distribution
do not depend on such a cut-off. 
Therefore, a Weibull distribution is more convenient 
than a Mittag-Leffler type one if one wishes to evaluate relevant 
statistics such as average waiting time 
in financial markets with long durations. 
On the other side, we find that the Gini index 
is rather independent of the cut-off parameter. 
Based on the above considerations, we propose a good 
candidate for describing the distribution of first-passage 
time in a market: The Weibull distribution 
with a power-law tail. This distribution compensates the gap between 
theoretical and empirical results 
much more efficiently than a simple Weibull distribution. 
We also give a useful formula to 
determine an optimal crossover point 
minimizing the difference between the empirical
average waiting time and the one predicted from
renewal theory.
Moreover, 
we discuss the limitation of 
our distributions 
by applying our distribution 
to the analysis of the BTP future and 
calculating 
the average waiting time. 
We find that our distribution 
is applicable as long as 
durations follow a Weibull-law for short times and 
do not have too heavy a tail. 
\end{abstract}

\end{frontmatter}

%%%%%%%%%%%%%%%%%%%%%%%%%%%%%%%%%%%%%%%%%%%%%%%%%%%%%%%%%%%%%%%%%%%
%%                      Introduction                             %%
%%%%%%%%%%%%%%%%%%%%%%%%%%%%%%%%%%%%%%%%%%%%%%%%%%%%%%%%%%%%%%%%%%%
\section{Introduction}
\label{intro}
%%%%%%%%%%%%%%%%%%%%%%%%%%%%%%%%%%%%%%%%%%%%%%%%%%%%%%%%%%%%%%%%%%
The distribution of time intervals between price changes 
gives us important pieces of information about the market \cite{Scalas}.
In particular, the fact that intertrade durations are not
exponentially distributed rules out the possibility
of using pure-jump L\'evy stochastic processes (i.e. compound
Poisson processes) as models for tick-by-tick
data. L\'evy processes have stationary and independent increments
and are Markovian and all these properties are a consequence
of exponentially distributed waiting times \cite{Scalas}. Other
models have been proposed such as non-homogeneous compound Poisson 
processes, GARCH-ACD models, continuous-time random
walks and semi-Markov processes \cite{hautsch,jenssen,engle,Scalas2,Scalas3,masoliver}.
 
Recently, various on-line trading services on the internet 
were established by several major banks. For instance, 
the Sony Bank uses a trading system in which 
foreign currency exchange rates change according to 
a first-passage process. 
Namely, the Sony Bank USD/JPY exchange rate is updated only 
when a reference market rate fluctuates by more than or equal to $0.1$ yen
\cite{SonyBank}. 
As a result, in the case of the Sony Bank rate, 
the average duration between price changes becomes longer, 
passing from $7$ seconds to $20$ minutes.
Automatic FOREX trading systems such as the one offered
by the Sony Bank are very popular in Japan where many investors
use a scheme called {\em carry trade} by borrowing money in a currency
with low interest rate and lending it in a currency offering higher
interest rates. As Japanese bond yields are low and US bonds offer
higher interest rates and are rated as safe financial instruments,
there is much trade in the USD/JPY market.
 
In this paper, we wish to compare the time structure
of the Sony bank trades with other markets
such as BTP futures (BTP is the middle and long term 
Italian Government bonds with fixed interest rates) once 
traded at LIFFE (LIFFE stands for London International 
Financial Futures and Options Exchange).
 
From the view-point of complex system engineering, 
a relevant quantity used to specify the stochastic process 
of the market rate is the average waiting time (a.k.a. residual life time) 
rather than the average duration. In a series of recent studies 
by the present authors, the average waiting time of the 
Sony Bank USD/JPY exchange rate was evaluated
under the assumption that the first-passage time (FPT) is a renewal
process whose distribution obeys a Weibull-law. 
We found that, counter-intuitively, the average waiting time 
of Sony Bank USD/JPY exchange rate is more than twice of the average duration 
\cite{InoueSazuka}. This fact is known as {\it inspection paradox}.
It means in general that the average of 
durations is shorter than the average 
waiting time. 
This fact is quite counter-intuitive 
because the customer checks the rate 
at the time between arbitrary 
consecutive rate changes. 
We shall explain the interpretation of this fact 
for the case in which durations follow 
the Weibull distribution. 

The Weibull distribution is often used for modelling intertrade durations in
financial markets \cite{hautsch,politi}. On the other side, 
the so-called Mittag-Leffler survival function has been also
proposed to represent the distribution of durations in several markets. 
For example, Mainardi et al. \cite{Raberto} showed that BTP future inter-trade durations are 
well-described by a survival function of Mittag-Leffler type. 
However, up to now, the Mittag-Leffler survival function has never 
been applied to evaluation of the average waiting time as it has
infinite moments of any integer order.

In this paper, we compare a Weibull distribution 
with a Mittag-Leffler type survival function in order to evaluate 
the average waiting time.
We give an analytical formula for the average waiting 
time under the assumption that the FPT distribution might 
be described by a Mittag-Leffler survival function. 
We find that the average waiting time diverges linearly 
with respect to a cut-off parameter $t_{\rm max}$. This fact tells us 
that it is hard to handle the Mittag-Leffler survival 
function to evaluate relevant statistics such as the average 
waiting time. 
We next evaluate the Gini 
index as another relevant statistic 
to check the 
usefulness of the Mittag-Leffler survival function. 

We also provide a good candidate for 
the description of the first-passage 
process of the market rates, namely, 
a Weibull distribution in which the behavior 
of the distribution changes from 
a Weibull-law to a power-law at 
some crossover point $t_{\times}$. 
We find that the average waiting time becomes much closer to the
empirical value for the Sony Bank 
USD/JPY exchange rate than for a pure Weibull distribution.
We also give a useful formula to 
determine the optimal crossover point 
in the sense that 
the gap of 
the average waiting 
time between the empirical and 
the proposed distributions 
is minimized for the crossover point. 
Moreover, 
we discuss the limitation of 
our distribution 
by applying our distribution 
to the analysis of the BTP future and 
calculating 
the average waiting time. 
We find that our distribution 
is applicable as long as 
duration 
follows a Weibull-law in short duration 
regime and 
does not have 
too heavy a tail. 

As mentioned above, in this paper, two sets of data
are used. The first set comes from the Sony bank and the
random variable analysed is a {\em first-passage time}, whereas
the second set is made up of future BTP prices traded at LIFFE
in 1997 for two different maturities: June and September. For these
data, the relevant random variable is an {\em intertrade duration}.
Both data sets have already been studied and extensively described in
previous papers 
(\cite{InoueSazuka,Raberto,Sazuka2,Sazuka,SazukaInoue,SazukaInoue2}). 
In both
cases, we assume that the empirical random variables are a realization
of a renewal process. A renewal process is a one-dimensional
point process where at times $T_0, T_1, \ldots, T_n, \ldots$
some event takes place, and the differences $\tau_i = T_i - T_{i-1}$
are independent and identically distributed (i.i.d.) random variables, so that
$T_n = \sum_{i=1}^{n} \tau_i$.: Therefore $T_n$ can be seen as a sum
of non-negative i.i.d. random variables, that is as an instance of random walk.
For the Sony bank data the incoming events are price changes due to
crossing the $\pm 0.1$ yen level around the current price, whereas in the
BTP-future case, the events are consecutive trades. Therefore, in the
Sony bank case, the waiting time is the residual life-time to next passage
and in the BTP-future case, the waiting time is the residual life-time to the next trade.

This paper is organized as follows. 
In the next section, we introduce both 
the Mittag-Leffler survival function and 
the Weibull distribution. Then, we 
discuss their properties in detail. 
In section 3, we evaluate the average waiting 
time for the Mittag-Leffler survival function. 
We find that the average waiting time 
diverges linearly as a function of the cut-off parameter 
$t_{\rm max}$. In section 4, 
we provide a theoretical formula of 
the Gini index for 
the Mittag-Leffler function and 
we check the usefulness 
by comparing 
the theoretical prediction with  
empirical data analysis 
for the BTP future. 
In section 5, we introduce 
a Weibull distribution with a power-law tail 
to compensate a small gap between the results of 
theoretical and empirical data analysis for 
the average waiting time. 
In the same section, 
we give an intuitive explanation 
for the non-monotonic behavior of 
the average waiting time 
corrected by means of the 
Weibull distribution with a power-law tail. 
From the observation, 
we obtain a useful formula 
to determine the optimal 
crossover point 
for which 
the gap between theoretical prediction 
and the empirical data analysis for 
the average waiting time 
is minimized. 
In section 6, 
we apply our distribution 
to the BTP future to 
check the limitation of 
our approach. 
In the final section 7, 
we summarize and discuss our results.  
%%%%%%%%%%%%%%%%%%%%%%%%%%%%%%%%%%%%%%%%%%%%%%%%%%%%%%%%%%%%%%%%%
\section{Mittag-Leffler survival function and Weibull distribution}
%%%%%%%%%%%%%%%%%%%%%%%%%%%%%%%%%%%%%%%%%%%%%%%%%%%%%%%%%%%%%%%%%%
For BTP-future data, the successive time intervals 
are reasonably described in terms of the Mittag-Leffler survival 
function \cite{Raberto}: 
%%%%%%%%%%%%%%%%%%%%%%
\begin{eqnarray}
E_{\beta}(-(t/t_{0})^{\beta}) & = & 
\sum_{n=0}^{\infty}
(-1)^{n} 
\frac{(t/t_{0})^{\beta n}}
{\Gamma (\beta n +1)}\,\,\,\,\,\,\,\,
(0 < \beta \leq 1)
\label{eq:Mittag}
\end{eqnarray}
%%%%%%
where $\Gamma (z)$ denotes the Gamma function;
we set the upper bound of the sum to a large value 
$n_{\rm max}$ for practical numerical calculations. 
%%%%%%%%%%%%
The above Mittag-Leffler 
survival function has asymptotic forms: 
$$E_{\beta}(-(t/t_{0})^{\beta}) 
\simeq {\exp}[
-(t/t_{0})^{\beta}/ \Gamma (1+\beta)]\,\,\,\,(t/t_{0} \to 0)$$ 
(stretched exponential) and 
$$E_{\beta}(-(t/t_{0})^{\beta}) \simeq 
(t/t_{0})^{-\beta}/ \Gamma (1-\beta)
\,\,\,\,(t/t_{0} \to \infty).$$ 
%%%%%%%%%%%%%%%%%%%%%%%%%%%%%%%%%
\begin{figure}
\begin{center}
\includegraphics[width=11cm]{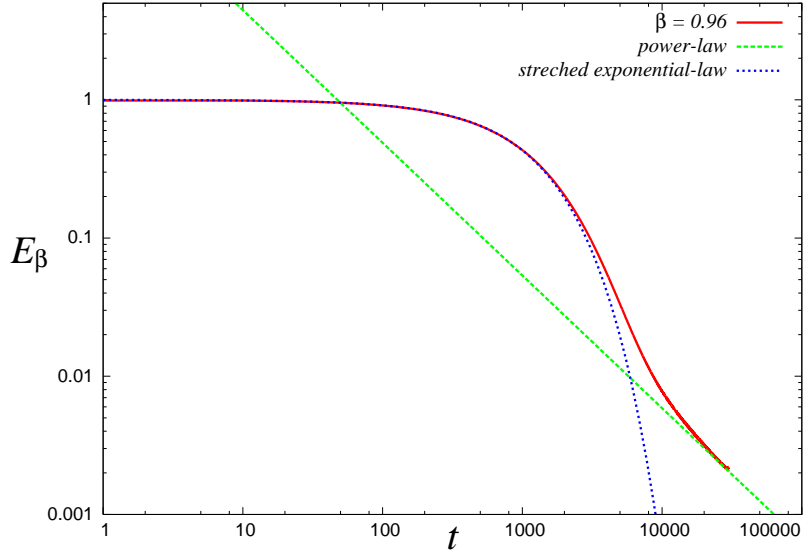}
\caption{\footnotesize 
The behavior of the 
Mittag-Leffler survival function (\ref{eq:Mittag}). 
The parameters are $\beta =0.96$ and $t_{0}=1200$. 
}
\label{fig:fg0}
\end{center}
\end{figure}
%%%%%%%%%%%%%%%%%%%
We illustrate these asymptotic forms 
in Fig. \ref{fig:fg0}. 
%%%%%%%%%%%
Then, the density function of the duration $t$ is given by
%%%%%%%%%%%
%%%%
\begin{eqnarray}
P_{ML} (t: t_0, \beta) & \equiv & 
-\frac{\partial E_{\beta}(-(t/t_{0})^{\beta})}
{\partial t} = 
\frac{1}{t_{0}}
\sum_{n=0}^{\infty}
(-1)^{n}
\frac{(t/t_{0})^{\beta n + \beta -1}}
{\Gamma (\beta n + \beta)}.
\label{eq:density_ML}
\end{eqnarray}
In the limiting case $\beta = 1$, the Mittag-Leffler distribution
coincides with the exponential distribution.
%%%%%%%%%%%%%%%%%
On the other hand, the so-called Weibull distribution has 
a probability density function given by 
%%%%%%
%%%%%%%%%%%%%%%%%%%%%%%%%%%%
\begin{eqnarray}
P_{W}(t: m,a) & = & 
m\,
\frac{t^{m-1}}{a}\,
{\exp}
\left(
-\frac{t^{m}}{a}
\right), 
\label{eq:Weibull}
\end{eqnarray}
%%%%%%%%%%%%%%%
and is a good approximation to the passage times for the Sony Bank USD/JPY exchange 
rate in a non-asymptotic regime $t \ll \infty$. It can be directly verified that the
Weibull distribution (\ref{eq:Weibull}) becomes 
an exponential distribution for $m=1$ and a Rayleigh 
distribution for $m=2$. 

For these two candidate distributions, we study a relevant statistic: the average waiting time, 
a quantity used in queueing theory,
which has been defined in the introduction as the residual life-time for a
renewal process. 
%%%%%%%%%%%%%%%%%%%%%%%%%%%%%%%%%%%%%%%%%%%%%%%%%%%%%%%%%%%%%%%%
\section{Divergence of the average waiting time for 
the Mittag-Leffler survival function}
\begin{figure}[ht]
\begin{center}
\includegraphics[width=11cm]{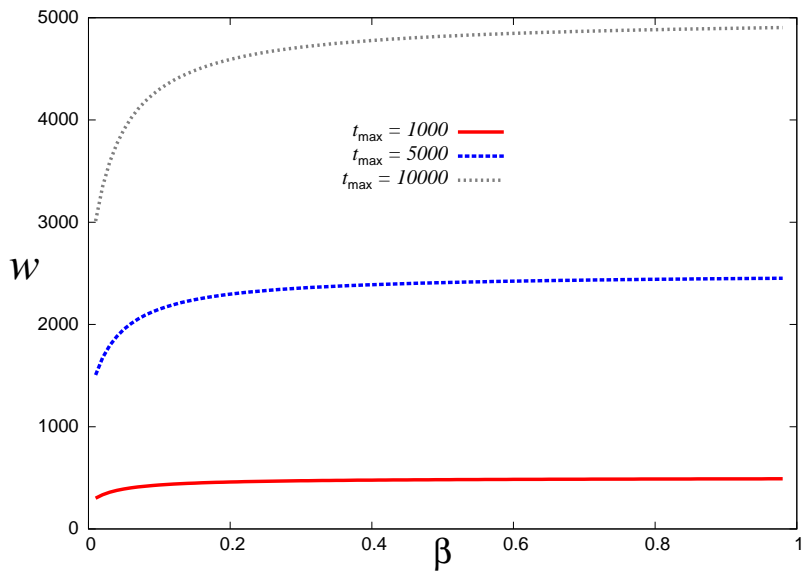}
\caption{\footnotesize 
The average waiting time 
$w$ by (\ref{eq:AWT2}) with 
$t_{0}^{-1}=1/12$.  
}
\label{fig:fg1}
\end{center}
\end{figure}

\begin{figure}[ht]
\begin{center}
\includegraphics[width=11cm]{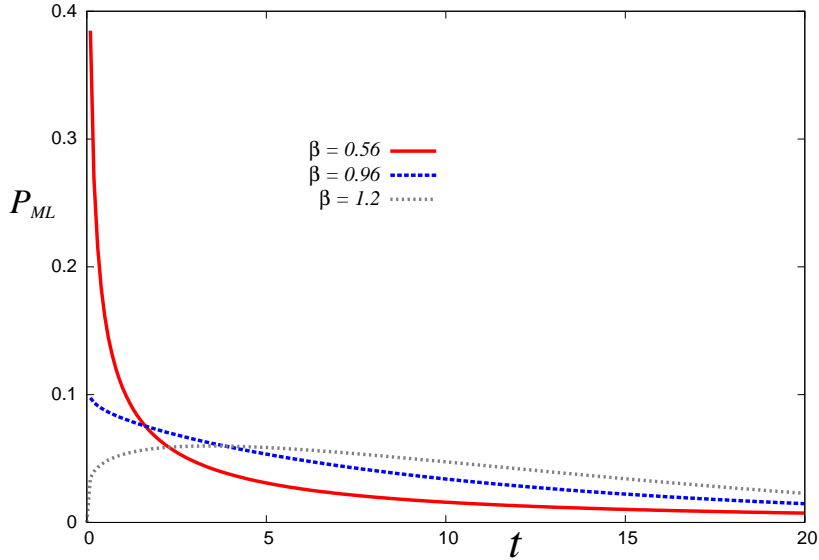}
\caption{\footnotesize 
The short-time-range 
behavior of the approximate  
probability density defined in eq. (\ref{approximant}) 
for several values of $\beta$. 
We set $t_{\times}=12$. 
}
\label{fig:fg12}
\end{center}
\end{figure}

%%%%%%%%%%%%%%%%%%%%%%%%%%%%%%%%%%%%%%%%%%%%%%%%%%%%%%%%%%%%%%%%
The first two moments of the Mittag-Leffler distribution diverge.
For the average of a random variable $t$,
we use the notation $\mathbb{E}(t)$.
It can be shown that also the residual life-time, defined as the
ratio of the first two moments of the distribution diverges.
One possibility is truncating the Mittag-Leffler distribution
at some time $t_{\mathrm{max}}$ and normalizing to
$\int_{0}^{t_{\mathrm{max}}} P_{ML} (t) \, dt$. This distribution
has finite moments of all orders and it turns out that the waiting time
$w=\mathbb{E}(t^{2})/2\mathbb{E}(t)$ is: 
%%%%%%%%
%%%%%%
\begin{eqnarray}
w (t_{0},t_{\rm max}: \beta) & = & 
\frac{t_{0}}{2} 
\frac{
\sum_{n=0}^{n}
(-1)^{n}
\frac{(t_{\rm max}/t_{0})^{\beta n + \beta +2}}
{(\beta n+\beta +2)\Gamma (\beta n +\beta)}
}
{
\sum_{n=0}^{n}
(-1)^{n}
\frac{(t_{\rm max}/t_{0})^{\beta n + \beta +1}}
{(\beta n+\beta +1)\Gamma (\beta n +\beta)}
}.
\label{eq:AWT2}
\end{eqnarray}
%%%%%%%%%%%
%%%%%%%%%%%%%%%%%%%%%
In Figure \ref{fig:fg1}, we plot the $w$ for several 
values of $t_{\rm max}$ with $t_{0}^{-1}=1/12$. 
%%%%%%%%%%%%%%%
%%%%%%%%%%%%%%%%%%%%%%%
%%%%%%%%%%%%%%%%%%%%%%%%%%%%%%
However, we should keep in mind that the above $w$ diverges as 
$t_{\rm max} \to \infty$. As we saw, the asymptotic form of 
the above density function is $\sim t^{-1-\beta}$ when 
$t/t_{0} \to \infty$. The divergence of the $w$ might come from 
only this power-law regime. Actually, we see this fact by 
evaluating the first two moments of the density function in the tail region. 
These two moments behave as $\mathbb{E}(t^{2}) 
\simeq 
\int_{0}^{t_{\rm max}} r^{-\beta-1} r^{2} dr =t_{\rm max}^{2-\beta}$ and 
$\mathbb{E}(t)\simeq 
\int_{0}^{t_{\rm max}} r^{-\beta -1} rdr=t_{\rm max}^{1-\beta}$ for 
(\ref{eq:AWT2}) as $t_{\rm max} \to \infty$. 
Thus, the average waiting time diverges linearly as a function of $t_{\rm max}$ 
as $w \simeq t_{\rm max}^{2-\beta}/t_{\rm max}^{1-\beta}=t_{\rm max}$. 
We can now define an {\em effective} probability density which approximates
the Mittag-Leffler distribution as follows:  
%%%%
\begin{eqnarray}
\label{approximant}
P_{ML}(t) & \simeq & 
\left\{
\begin{array}{cc}
P_{S}(t) & (t \leq t_{\times}) \\
t_{\times}^{\beta +1} P_{S}(t_{\times})\, t^{-1-\beta} & (t > t_{\times}) 
\end{array}
\right.
\end{eqnarray}
%%%%%
where 
$P_{S}(t)$ is a stretched exponential 
distribution. 
With this approximation, one gets 
%%%%%
\begin{eqnarray}
w(t_{\times},t_{\rm max}:\beta) & \simeq & 
\frac{\int_{0}^{t_{\times}}
t^{2}P_{S}(t) dt + 
t_{\times}^{\beta +1} P_{S}(t_{\times}) \int_{t_{\times}}^{t_{\rm max}}
t^{1-\beta} dt}
{2\int_{0}^{t_{\times}}
t P_{S}(t) dt + 2 
t_{\times}^{\beta +1} P_{S}(t_{\times}) \int_{t_{\times}}^{t_{\rm max}}
t^{-\beta}dt} \nonumber \\
\mbox{} & \simeq & 
\frac{\int_{0}^{t_{\times}}
t^{2}P_{S}(t) dt + 
t_{\times}^{\beta +1} P_{S}(t_{\times})\, t_{\rm max}^{2-\beta} +{\mathcal O}(1)}
{2\int_{0}^{t_{0}}
t P_{S}(t) dt + 2 
t_{\times}^{\beta +1} P_{S}(t_{\times}) \, t_{\rm max}^{1-\beta} +{\mathcal O}(1)} \nonumber \\
\mbox{} & \simeq & 
\left\{
\begin{array}{lr}
w_{S}(t,t_{\times}:\beta) + {\mathcal O}(1) & (\beta \geq 2) \\
%%%
t_{\rm max}^{2-\beta} & (1 \leq \beta < 2) \\
%%%%
t_{\rm max} & ( 0 < \beta < 1) 
\end{array}
\right..
\end{eqnarray}
%%%%%
Notice that, for this approximant of the Mittag-Leffler function, it is meaningful
to consider $\beta > 1$, as one can build a legitimate probability
density (a non-negative function of positive reals normalized to 1) for any $\beta >0$.
However, the Mittag-Leffler function is no longer a legitimate survival function
for $\beta >1$ as it assumes negative values. For $\beta > 2$ the approximant function
has finite first and second moment and also the waiting time
$w_{S}(t,t_{\times}:\beta)$ has a 
finite value. 
Thus, 
if we were restricted to choose the 
parameter $\beta$ within 
the range 
$0 < \beta < 1$, 
the average waiting time 
$w$ would diverge as $\sim t_{\rm max}$.  
If we could choose 
$\beta>2$, 
we would obtain a finite 
value of the average waiting time, 
however, for $\beta >1$, 
the approximate probability density
has a maximum within the range $t<\infty$. 
In Fig. \ref{fig:fg12}, the behavior of the approximate 
density is shown  for several values of 
the parameter $\beta$ in 
the short time regime. 
From this figure, we find that 
the maximum appears for $\beta >1$. 
%%%%%%%%%%%%%%%
%%%%%%%%%%%%%%%%%%%%%%%
This behavior is quite 
different from the 
empirical probability density function. 
Moreover, as mentioned above, 
for $\beta > 1$, 
the Mittag-Leffler distribution cannot
be used. 
Thus, we should use a truncated Mittag-Leffler distribution and include a finite upper bound of 
the integral with respect to $t$, namely, {\it the maximum value 
of the duration} or {\it the cut-off parameter} $t_{\rm max}$.  

In the latter case, we have to face the following problem. Namely, how do we determine $t_{\rm max}$ 
to obtain a reasonable $w$ that is consistent with the result 
obtained from the empirical data analysis? 
Unfortunately, the estimate of $w$ also depends on a second
parameter, the crossover point $t_{\times}$ 
at which the density function changes its shape from a stretched 
exponential-law to a power-law. 
If  $t_{\times}$ is close to $t_{\rm max}$, the value of $w$ is 
not sensitive to the value of 
 $t_{\rm max}$; however, if  
$t_{\times}$ is far from  $t_{\rm max}$, 
 $w$ does depend on the value of  $t_{\rm max}$ because the 
integral of the power-law tail becomes dominant. These considerations
lead to the conclusion 
that the Mittag-Leffler function is hard 
to use in order to evaluate the average waiting time for the market rates 
with a relatively long duration such as 
the Sony Bank USD/JPY exchange rate.
%%%%%%%%%%%%%%%%%%%%%%%%%%%%%%%%%%%%%%%%%%%%%%%%%%%%%%%%%%%%
\section{The Gini index}
%%%%%%%%%%%%%%%%%%%%%%%%%%%%%%%%%%%%%%%%%%%%%%%%%%%%%%%%%%%%
Another relevant statistic to specify the 
market rate with a long duration is 
the so-called Gini index, which denotes 
the inequality of the durations used this paper. 
In other words, fluctuation level in duration lengths
can be simply described in terms of the Gini index.
For a Weibull distribution, 
it was shown that 
the Gini coefficients given by both 
analytical prediction and empirical evidence 
coincide \cite{SazukaInoue2}. 
However, 
for the Mittag-Leffler 
survival 
function, 
it is not clear whether 
the analytical 
prediction of 
the Gini 
index is close to 
the corresponding empirical evidence 
due to the tail-effect discussed in the previous section. 
Here, we study this issue. 
%%%%%%%%%%%%%%%%%%%%%%%%%%%%%%%%%%%%%%%%%%%%%%
\subsection{Analytical evaluation}
\begin{figure}[ht]
\begin{center}
\includegraphics[width=11cm]{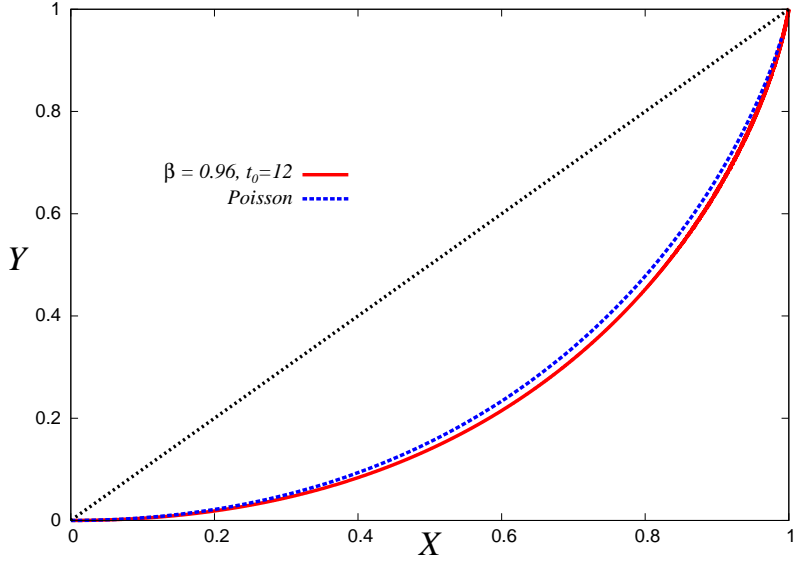}
\caption{\footnotesize 
The Lorentz curve for 
the Mittag-Leffler survival function. 
We set $t_{0}=12$ and $\beta=0.96$. 
We set the effective upper-bound 
of the integral $r_{\rm max}=100$.
We also plot the 
Lorentz curve for 
the Poisson process, 
namely, 
for the 
exponential duration for 
which 
the curve is 
written explicitly 
$Y=X+(1-X)\log (1-X)$. 
$Y=X$ is the perfect equality line. 
}
\label{fig:fg13}
\end{center}
\end{figure}

\begin{figure}[ht]
\begin{center}
\includegraphics[width=11cm]{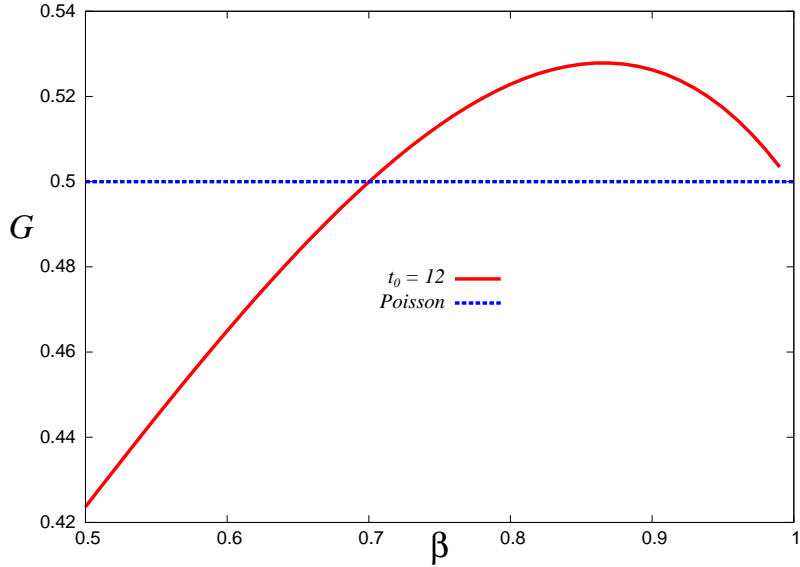}
\caption{\footnotesize 
The Gini index as 
a function of $\beta$ for 
the truncated Mittag-Leffler survival function. 
We set $t_{0}=12$. 
The constant horizontal line 
$G=1/2$ corresponds to 
the Gini index for 
the exponential duration. 
We set the effective upper-bound 
of the integral $r_{\rm max}=100$. 
}
\label{fig:fg14}
\end{center}
\end{figure}

%%%%%%%%%%%%%%%%%%%%%%%%%%%%%%%%%%%%%%%%%%%%%%%%%%%%
The Gini index $G$ is defined as 
the area between the Lorentz curve defined below: 
$(X(r),Y(r))$ $r \in [0,\infty]$ and 
the line $Y=X$ corresponding to perfect equality, 
namely, 
%%%%%%%%%%%%%%%%%%%
\begin{eqnarray}
G & = & 
\int_{0}^{1}
(X-Y)dX = 
\int_{0}^{\infty}
\{X(r)-Y(r)\} \frac{dX}{dr} \cdot dr. 
\end{eqnarray}
%%%%%%%%%
For the truncated Mittag-Leffler distribution, 
the Lorentz curve can be 
calculated as 
%%%%%%%%%
\begin{eqnarray}
X(r) & = & 
\frac{\int_{0}^{r} dt P_{ML}(t: t_0, \beta)}
{\int_{0}^{r_{\rm max}} dt P_{ML}(t: t_0, \beta)} \simeq 
1- 
\sum_{n=0}^{\infty}(-1)^{n}
\frac{(r/t_{0})^{\beta n}}
{\Gamma (\beta n +1)} \\
%%%%%
Y(r) & = &  
\frac{\int_{0}^{r} dt t P_{ML}(t: t_0, \beta)}
{\int_{0}^{r_{\rm max}} dt t P_{ML}(t: t_0, \beta)} \nonumber \\
\mbox{} & = & 
\frac{
-r \sum_{n=0}^{\infty}
(-1)^{n}
\frac{(r/t_{0})^{\beta n}}
{\Gamma (\beta n +1)}
+ 
r 
\sum_{n=0}^{\infty}
(-1)^{n}
\frac{(r/t_{0})^{\beta n}}
{(\beta n + 1)\Gamma (\beta n + 1)}
}
{
-r_{\rm max} \sum_{n=0}^{\infty}
(-1)^{n}
\frac{(r_{\rm max}/t_{0})^{\beta n}}
{\Gamma (\beta n +1)}
+ 
r_{\rm max} 
\sum_{n=0}^{\infty}
(-1)^{n}
\frac{(r_{\rm max}/t_{0})^{\beta n}}
{(\beta n + 1)\Gamma (\beta n + 1)}
} \nonumber \\
\mbox{} & = & 
\frac{r \sum_{n=0}^{\infty}
n(-1)^{n}
\frac{(r/t_{0})^{\beta n}}{(\beta n + 1) \Gamma (\beta n +1)}
}
{r_{\rm max} \sum_{n=0}^{\infty}
n(-1)^{n}
\frac{(r_{\rm max}/t_{0})^{\beta n}}{(\beta n + 1) \Gamma (\beta n +1)}
}.
\end{eqnarray}
%%%%%%%%%%%%%
In Fig. \ref{fig:fg13}, 
we plot the Lorentz curve for 
the parameters $\beta = 0.96$ and 
$t_{0}=12$ (according to 
reference \cite{Raberto}), but with an effective
upper bound of the integral set at $r_{\rm max}=100$. 
%%%%%%
In the same figure, 
we show the Lorentz curve for 
the 
the Poisson process, 
namely, 
for the 
exponential duration for 
which 
the curve can be 
written explicitly 
$Y=X+(1-X)\log (1-X)$. 
From this figure, 
one can see that 
the area between 
the Lorentz curve for the Mittag-Leffler 
and the perfect 
equality line $Y=X$ is 
larger than the area between 
the Lorentz curve for 
the Poisson 
process and $Y=X$. 
This means that 
the durations 
generated 
from the Mittag-Leffler survival 
function is more 
biased than that of the Poisson process. 
This fact can be justified by directly calculating
Gini's index. 
%%%%%%%%%%%%%%%
%%%%%%%%%%%%%%%%%%%%%%%
For the Lorentz curve of the truncated Mittag-Leffler distribution, 
$G$ is written as follows:
%%%%%%%%%
\begin{eqnarray}
G & = & 2\int_{0}^{r_{\rm max}}
dr 
{\Biggr \{}
1- \sum_{n=0}^{\infty} (-1)^{n}
\frac{(r/t_{0})^{\beta n}}
{\Gamma (\beta n +1)} - 
\frac{r\sum_{n=0}^{\infty}
(-1)^{n} 
\frac{n (r/t_{0})^{\beta n}}
{(\beta n+1) \Gamma (\beta n+1)}
}
{r_{\rm max}\sum_{n=0}^{\infty}
(-1)^{n} 
\frac{n (r_{\rm max}/t_{0})^{\beta n}}
{(\beta n+1) \Gamma (\beta n+1)}
}
{\Biggr \}} \nonumber \\
\mbox{} & \times & 
\frac{1}{t_{0}}
\sum_{n=0}^{\infty}
(-1)^{n}
\frac{(r/t_{0})^{\beta n +\beta -1}}
{\Gamma (\beta n + \beta)} \nonumber \\
\mbox{} & = & 
\frac{2r_{\rm max}}{\beta t_{0}}
\sum_{n=0}^{\infty}
\frac{(-1)^{n}(r_{\rm max}/t_{0})^{\beta n + \beta -1}}
{(n + 1) 
\Gamma (\beta n + \beta)} \nonumber \\
\mbox{} & - & 
\frac{2r_{\rm max}}{\beta t_{0}}
\sum_{n=0}^{\infty}
\sum_{l=0}^{\infty}
\frac{(-1)^{n+l} (r_{\rm max}/t_{0})^{\beta (n+l) + \beta -1}}
{(n+l+1)\Gamma (\beta n+1)
\Gamma (\beta l + \beta)} \nonumber \\
\mbox{} & - & 
\frac{2r_{\rm max}}
{t_{0}}
\frac{
\sum_{n=0}^{\infty}
\sum_{l=0}^{\infty}
\frac{n 
(-1)^{n+l}(r_{\rm max}/t_{0})^{\beta (n+l) + \beta -1}}
{(\beta (n+l) +\beta +1)(\beta n+1)
\Gamma (\beta n + 1)
\Gamma (\beta l +\beta)}
}
{\sum_{n=0}^{\infty}
(-1)^{n} 
\frac{n (r_{\rm max}/t_{0})^{\beta n}}
{(\beta n+1) \Gamma (\beta n+1)}
}.
\end{eqnarray}
%%%%%%%%%%%%%
We plot the Gini index 
$G$ as a function of $\beta$ for 
$t_{0}=12$ and $r_{\rm max} = 100$
in Fig. \ref{fig:fg14}. 
%%%%%%%%%%%%%%%%%%%%%%%%%%%%%%%%%%
%%%%%%%%%%%%%%%
%%%%%%%%%%%%%%%%%%%%%%
From this figure, we find that 
the Gini index for $\beta=0.96$ 
is $G=0.51$ and 
$G$ approaches 
$1/2$ which is the Gini index for 
the exponential duration. 
For 
both the Lorentz curve and 
the Gini index, 
we set the effective upper-bound of 
the integral as $r_{\rm max}=100$, however, 
we find that 
this statistic is 
free from the kind of divergence affecting 
the average waiting time $w$ due to 
the upper-bound. 
%%%%%%%%%%%%%%%%%%%%%%%%%%%%%%%%%%%%%%%%%%%%%%%%%%%%%%%%%%%%
\subsection{Empirical data analysis}
%%%%%%%%%%%%%%%%%%%%%%%%%%%%%%%%%%%%%%%%%%%%%%%%%%%%%%%%%%%%
Based on the method 
proposed in \cite{SazukaInoue}, 
we obtain 
$G=0.59$ for 
the BTP future with maturity
June and 
$G=0.57$ for 
the BTP future with maturity 
September, whereas the theoretical prediction
obtained in the previous section is 0.51. 
From these results, 
we find a manifest gap between 
the theory and 
empirical data analysis, 
however, 
this gap is 
relatively small 
in 
comparison 
with 
the gap for the average waiting 
time as we shall see later. 
%%%%%%%%%%%%%%%%%%%%%%%%%%%%%%%%%%%%%%%%%%%%%%%%%%%%%%%%%%%%%
\section{A Weibull distribution with a power-law tail}
\begin{figure}
\begin{center}
\includegraphics[width=11cm]{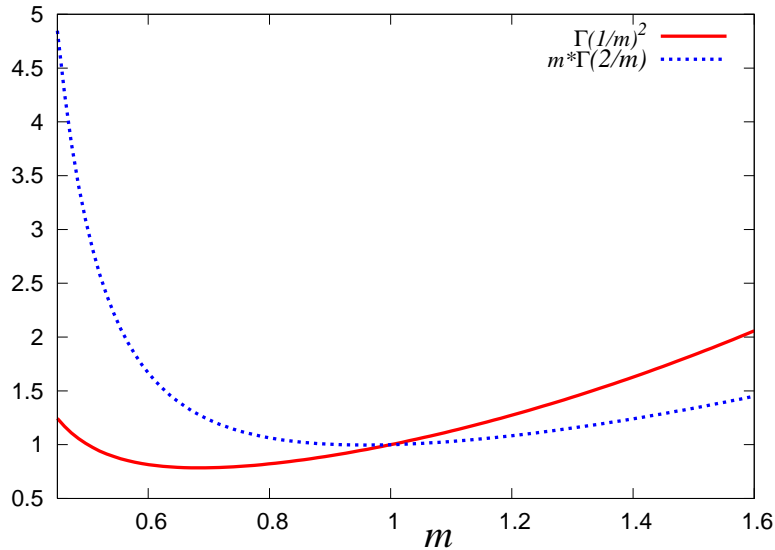}
\caption{\footnotesize 
$l_{1} \equiv {\Gamma(1/m)}^2$ 
and $l_{2} \equiv m\Gamma(2/m)$ as a function of 
Weibull parameter $m$.
At the intersection of both lines 
for $m=1$, 
the average waiting time for the 
Weibull distribution is equal to the average duration.
For $m>1$, the average waiting time 
is longer than the average duration, 
whereas for $m <1$, 
the so-called {\it inspection paradox} 
takes place \cite{InoueSazuka}. 
}
\label{fig:fg00}
\end{center}
\end{figure}

%%%%%%%%%%%%%%%%%%%%%%%%%%%%%%%%%%%%%%%%%%%%%%%%%%%%%%%%%%%%%
In previous studies, we found that a Weibull distribution 
is a good candidate to describe the Sony Bank USD/JPY 
exchange rate time statistic \cite{Sazuka}. The average waiting time was also 
evaluated to investigate to what extent the Sony Bank rate is 
well-explained by the Weibull distribution \cite{InoueSazuka,SazukaInoue}. 
We also found that the empirical result of 
the waiting time of Sony Bank USD/JPY exchange rate ($\sim 49.19$ [min]) 
is more than twice of the average duration ($\sim 20.52$ [min]).
The situation is known as inspection paradox 
as discussed in the introduction.
For the Weibull distribution,
the paradox occurs when the Weibull parameter 
satisfies $m<1$
as shown in Fig. \ref{fig:fg00}. 
In this plot, we 
used the fact that 
$\mathbb{E}(t)=a^{1/m} (1/m)\Gamma (1/m), 
w=a^{1/m} \Gamma (2/m)/\Gamma (1/m)$ for 
a Weibull distribution (\ref{eq:Weibull}) and 
the condition $\mathbb{E}(t)=w$ require 
$l_{1} \equiv \{\Gamma (1/m)\}^{2} = 
m\Gamma (2/m) \equiv l_{2}$. The solution of 
this equation $l_{1}=l_{2}$
gives $m=1$, and $m <1$ for $l_{1} > l_{2}$ means 
$\mathbb{E}(t) < w$, vice versa \cite{InoueSazuka}. 
This fact is intuitively understood as follows. 
When the parameter $m$ is smaller than $1$, 
the bias of the duration is larger than that of 
the exponential distribution. 
As a result, the chance for customers to 
check the rate within large intervals between 
consecutive price changes 
is more frequent than the chance they check the rate 
within short intervals. 
Then, the average waiting time could become 
longer than the average duration. 

The bias of the duration for 
the Weibull distribution with $m<1$ 
is directly confirmed by means of the Gini index. 
It was shown that the analytical prediction of the Gini 
index calculated for a Weibull distribution is in good agreement 
with the value obtained from the empirical data of the Sony Bank 
rate \cite{SazukaInoue2}. 

However, there exists a significant small gap between the 
theoretical prediction ($\sim 44.62$ [min]) and the empirical 
result for $w$ ($\sim 49.19$ [min]). 
%%%%%%%%%%%%%%%%%%%%%%%%%%%%%%%%%
%%%%%%%%%%%%%%%%%%%

In this section, we consider to what extent the average 
waiting time can be modified by taking into account a power-law 
behavior for the tail in the FPT distribution. 
In our previous paper, we assumed that the FPT of the Sony Bank 
rate might obey a pure Weibull distribution (\ref{eq:Weibull}). 
However, several empirical data analysis have shown that the 
shape of the FPT distribution changes from a pure Weibull-law 
to a power-law at some crossover point $t_{\times}$.  

\begin{figure}[ht]
\begin{center}
\includegraphics[width=11cm]{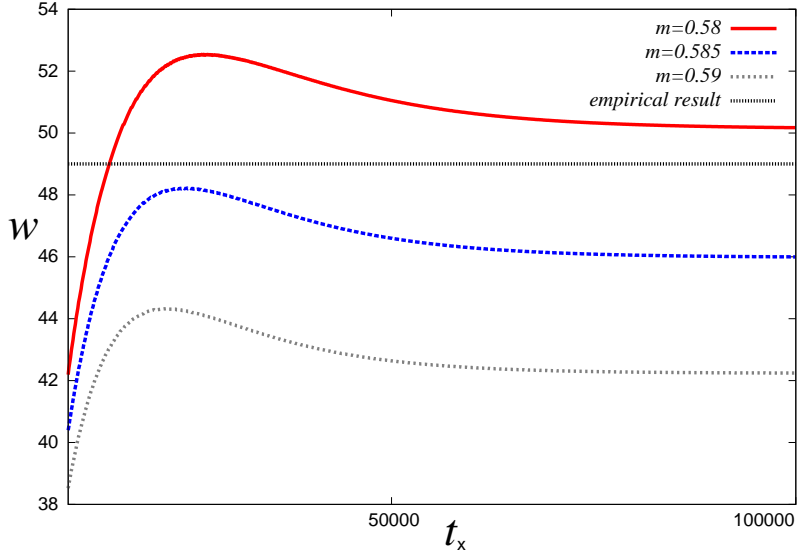}
\end{center}
\caption{\footnotesize 
The average waiting time $w$ as a function of $t_{\times}$. 
We set $\gamma = 4.67, a=49.63$. 
For two cases of the choice for $m$, namely, 
for $m=0.58, 0.585$ and 
$0.59$, 
the $w$ is plotted.}
\label{fig:fg3}
\end{figure}

Therefore, here it is natural to assume that the FPT distribution 
should be modified as follows: 
%%%%%%%%%%
\begin{eqnarray}
P_{\overline{W}}(t: m,a,\gamma,t_{\times}) & = & 
\left\{
\begin{array}{lr}
\frac{mt^{m-1}}{a}\,
{\exp}
\left(
{-\frac{t^{m}}{a}}
\right) & 
(t < t_{\times}) \\
\lambda\, t^{-\gamma} & 
(t > t_{\times}) 
\end{array}
\right.
\end{eqnarray}
%%%%%%%
Under the assumption of continuity at $t_{\times}$, 
the condition 
%%%%
\begin{eqnarray}
t_{\times}^{-\gamma}\, \lambda = 
\left( \frac{mt_{\times}^{m-1}}{a} \right)\,
{\exp}(-t_{\times}^{m}/a)
\end{eqnarray}
%%%%
is required. 
This condition determines the parameter $\lambda$ as  
%%%%%%%%%%%
\begin{eqnarray}
\lambda & = & 
\frac{mt_{\times}^{m+\gamma-1}}{a}
\,{\exp}
\left(
{-\frac{t_{\times}^{m}}{a}}
\right). 
\end{eqnarray}
%%%%%%%%%%%%%%%
Thus, the modified FPT distribution is given by 
%%%%%%%%%%%%%%%%%%%%%%%%%%
\begin{eqnarray}
P_{\overline{W}}(t: m,a,\gamma,t_{\times}) & = & 
\left\{
\begin{array}{lr}
\frac{mt^{m-1}}{a}\,
{\exp} \left(
{-\frac{t^{m}}{a}}
\right) & 
(t < t_{\times}) \\
\frac{mt_{\times}^{m+\gamma-1}}{a}\, 
{\exp} \left(
{-\frac{t_{\times}^{m}}{a}}
\right) 
 t^{-\gamma} 
& 
(t > t_{\times}) 
\end{array}
\right.
\label{eq:conti_cond}
\end{eqnarray}
%%%%%%%%%%%%%%%%%%%%%%%%%%%%%%%
%%%%%%%%%%%%%%%%%%%%%%%%%%%%%%%
From the FPT distribution, we have the average waiting time $w$ 
from the renewal-reward theorem as follows. 
%%%%%%%
\begin{eqnarray}
w (t_{\times}: m,a,\gamma) & = &  
\frac{
\frac{a^{1/m}}{m}\Gamma
\left(
\frac{1}{m}
\right) 
B
\left(
\frac{1}{m}+1, 
\frac{t_{\times}^{m}}{a}
\right) 
+ 
\frac{m t_{\times}^{m+1}}{a (\gamma-2)}\,
{\exp}
\left(
{-\frac{t_{\times}^{m}}{a}}
\right)
}
{
\frac{2a^{2/m}}
{m}
\Gamma 
\left(
\frac{2}{m}
\right) 
B
\left(
\frac{2}{m}+1,
\frac{t_{\times}^{m}}{a}
\right) + 
\frac{mt_{\times}^{m+2}}{a (\gamma-3)}\,
{\exp}
\left(
{-\frac{t_{\times}^{m}}{a}}
\right)
}
\label{eq:formulaW}
\end{eqnarray}
%%%%%%%%%%%%%%%%%%%
where $B(a,x)$ denotes the following incomplete Gamma function: 
%%%%%%%
\begin{eqnarray}
B(a,x) & = & 
\frac{1}{
\Gamma (a)}
\int_{0}^{x}
t^{a-1}
{\rm e}^{-t} dt. 
\end{eqnarray}
%%%%%%%%%%%%%%%%%%%%%%
%%%%%%%%%%%%%%%%%%%%%%%%%%
The next problem is how to choose the parameters 
$\gamma, t_{\times}, m$ and $a$. 
Fortunately, we know these parameters from empirical 
data analysis \cite{Sazuka,SazukaInoue}. 
Substituting those parameters 
$\gamma = 4.67, 
m=0.585$ and 
$a=49.63$ into 
our formula (\ref{eq:formulaW}), 
we evaluate the average waiting time $w$ 
as a function of the crossover point $t_{\times}$. 
The result is plotted in Figure \ref{fig:fg3}. 
%%%%%%%%%%%%%%%%%%%%%%%%%%%%%%%%%%%%%%%%%%%%%%%%%%
%%%%%%%%%%%%%%%%%%%%%%%%%
In this figure, we present the average waiting time 
for three slightly different cases of $m$, namely, 
$m=0.58,0.585$ and $0.59$. 
From the empirical data analysis, we have $t_{\times} 
\simeq 18000$ [{\rm s}]. 
Therefore, we conclude that for $m=0.585$, the waiting time is 
estimated as $w=45.66$ [min]. 
This value is much closer to the sampling value $w=49.19$ [min] 
than the value obtained under the assumption of a pure Weibull 
distribution ($w=44.62$ [min]). 
Therefore, we conclude that a correction by taking into account 
the tail behavior of the 
Weibull distribution points into 
the right direction for estimating the average waiting time. 

The remaining 
gap $\Delta w = 49.19-45.66 =3.53$ [min] 
might be due to 
a rough estimation of the 
crossover point $t_{\times}$. 
In the next section, 
we propose a systematic 
procedure to determine the appropriate 
crossover point $t_{\times}$ so as to 
minimize the gap $\Delta w$ 
by considering 
the non-monotonic behavior of 
the average waiting time 
$w$ with respect to $t_{\times}$ as 
shown in Fig. \ref{fig:fg3}.
%%%%%%%%%%%%%%%%%%%%%%%%%%%%%%%%%%%%%%%%%%%%%%%%%%%%%%%%%%%%%
\subsection{Intuitive explanation of the non-monotonic behavior}
\begin{figure}[ht]
\includegraphics[width=7.5cm]{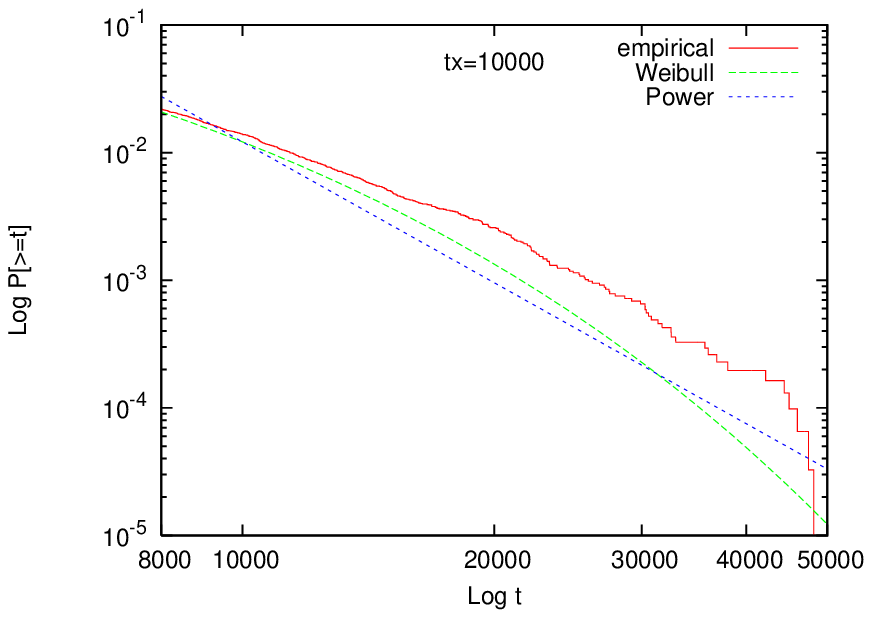}
\includegraphics[width=7.5cm]{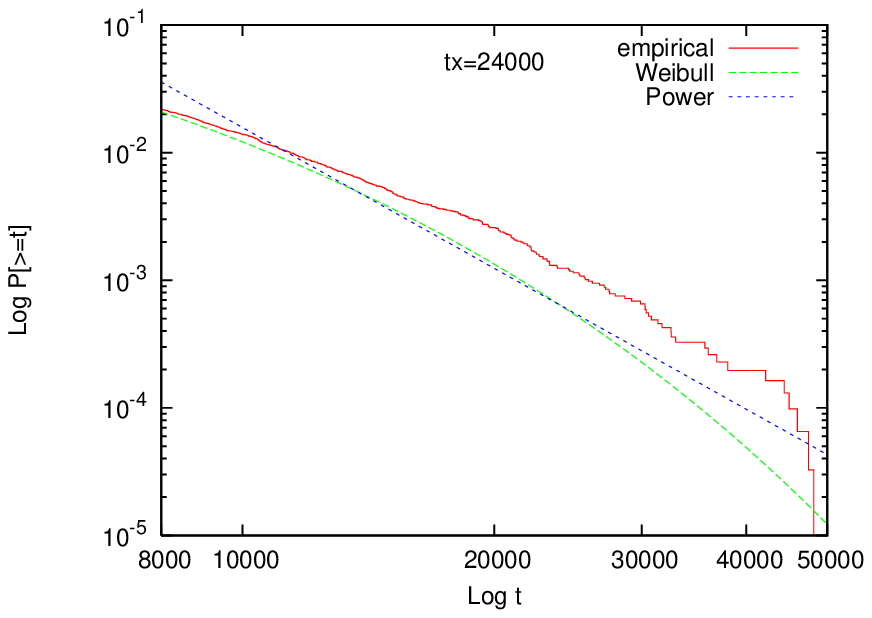}
\begin{center}
\includegraphics[width=7.5cm]{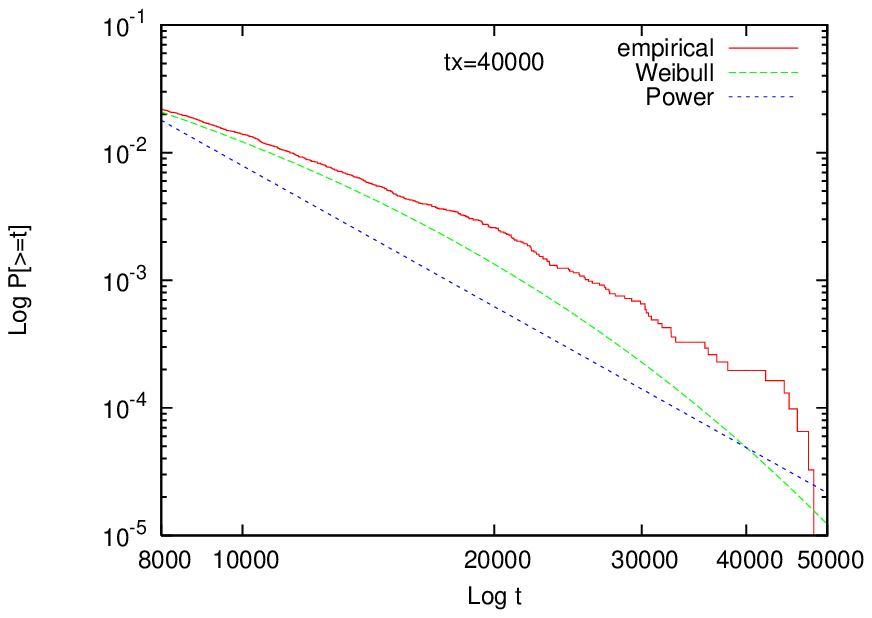}
\end{center}
\caption{\footnotesize 
The empirical distribution 
calculated from the Sony Bank rate 
and Weibull distribution 
with a power-law tail for 
$t_{\times} =10000$ [{\rm s}] (upper left), 
$24000$ [{\rm s}] (upper right) and 
$40000$ [{\rm s}].
}
\label{fig:fgA}
\end{figure}

%%%%%%%%%%%%%%%%%%%%%%%%%%%%%%%%%%%%%%%%%%%%%%%%%%%%%%%%%%%%%
From Fig. \ref{fig:fg3}, we find a non-monotonic behavior in 
the curve of the average waiting time as a function of $t_{\times}$. 
The intuitive explanation is given as follows. 
%%%%%
In Fig. \ref{fig:fgA}, we show the 
Log-Log plot of the survival function 
of the Weibull, the power-law and the empirical data 
for the regime $t >8000$ [{\rm s}]. In this figure, 
we set the crossover point 
$t_{\times}=10000$ [{\rm s}] to determine the normalization 
constant for the power-law distribution. 
Then, we find that there exists another 
intersection between the Weibull and the power-law distributions 
at $t \simeq 30000$ [{\rm s}]. 
As the results, we obtain 
two distinct areas which are surrounded 
by the two lines, namely, 
the Weibull and the power-law distributions. 
Let us call these two 
areas as 
$\mathcal{A}_{1}$ (the left part) and 
$\mathcal{A}_{2}$ (the right part), respectively. 
It should be noted that 
the difference $\epsilon$ between 
the empirical distribution 
and the Weibull distribution 
with a power-law tail 
is proportional to the  
difference of these two areas, 
that is, 
%%%%%%%%%
\begin{eqnarray}
\epsilon & \propto & 
\mathcal{A}_{1} -\mathcal{A}_{2}
\label{eq:def_area}. 
\end{eqnarray}
%%%%%%%
We should bear in mind that 
the above difference is 
dependent on the choice of the 
crossover point $t_{\times}$. 

In the following, 
we shall show that 
there exists an optimal 
crossover point $t_{\times}=C$ 
at which the difference $\epsilon$ is minimized. 
%%%%%%%%%%
\begin{itemize}
\item
%%%%%%%%%%%%%%%%%%%%%%%%%%%%%%%%%%
$10000 < t_{\times} < C$ \\
%%%%%%%%%%%%%%%%%%%%%%%%%%%%%%%%%%%%
For this case, 
as shown in the upper left of 
Fig. \ref{fig:fgA}, the curve of the 
power-law distribution 
goes up as the $t_{\times}$ increases. As the result, 
the area $\mathcal{A}_{1}$ decreases, whereas 
the area $\mathcal{A}_{2}$ increases. 
Then, 
the difference $\epsilon$, 
namely, 
the gap between 
the empirical distribution and 
the Weibull distribution with a power-law tail decreases. 
%%%%%%%%
\item
%%%%%%%%%%%%%%%%%%%%%%%%%%%%%%%%%%
$t_{\times}=C$ \\
%%%%%%%%%%%%%%%%%%%%%%%%%%%%%%%%%%
For this case, 
as shown in the upper right of 
Fig. \ref{fig:fgA}, 
the area $\mathcal{A}_{1}$ vanishes 
and the two distinct lines are 
degenerated to a single curve at $t_{\times}=C$. 
Then, the difference 
$\epsilon$ is minimized and 
the averaged waiting time with true 
parameters obtained by empirical data analysis 
takes its maximum. 
%%%%%%%
\item
%%%%%%%%%%%%%%%%%%%%%%%%%%%%%%%%%%%
$t_{\times} >C$ \\
%%%%%%%%%%%%%%%%%%%%%%%%%%%%%%%%%%%%%%
For this case, 
as shown in the lower Fig. \ref{fig:fgA},  
the curve of the power-law distribution 
goes down further and the single intersection 
at $t_{\times}=C$ moves to 
the right. As the result, 
the area $\mathcal{A}_{1}$ 
increases with the decreasing of 
the area $\mathcal{A}_{2}$. 
Thus,  from the definition of 
the difference (\ref{eq:def_area}), 
we find that for this case, 
the gap between the 
empirical distribution and 
the Weibull distribution 
with a power-law tail 
starts to increase again. 
%%%%%%%%%%%%
\end{itemize}
%%%%%
%%%%%%%%%%%%%%%%%%%%%%%
%%%%%%%%%%%%%%%%%%%%%%%%%

From the above observation, 
we conclude that 
the predicted average waiting time 
takes its maximum at $t_{\times}=C$. 

Taking into account the above fact, we 
might determine the 
optimal crossover point $t_{\times}^{*}$ for 
which the gap between 
the average waiting time 
for the empirical data and 
for the Weibull distribution 
with a power-law tail is minimized. 
In the next subsection, we 
shall discuss this issue. 
%%%%%%%%%%%%%%%%%%%%%%%%%%%%%%%%%%%%%%%%%%%%%%%%%%%%%%%%%%%%%%%%
\subsection{Determination of 
the optimal crossover point}
%%%%%%%%%%%%%%%%%%%%%%%%%%%%%%%%%%%%%%%%%%%%%%%%%%%%%%%%%%%%%%%
Let us consider the case $m=0.585$ in 
Fig. \ref{fig:fg3} which was evaluated from 
the Sony Bank rate by using the Weibull paper analysis. 
The error due to the wrong estimation for the true 
FPT distribution $P_{T}(t)$ can be divided into two parts, 
namely, the difference between the true (empirical) distribution 
and the Weibull distribution: 
$\varepsilon_{W}$, and the difference 
between the true distribution 
and the power-law distribution: 
$\varepsilon_{Power}$. 
Thus, the total difference $\varepsilon$ 
is written in terms of the area between the true 
curve $P_{T}(t)$, which is evaluated from 
empirical data analysis, and the Weibull distribution 
with a power-law tail. 
Then, we have the area
%%%%%
\begin{eqnarray}
\varepsilon & = & 
\int_{0}^{t_{\times}}
|P_{T}(t) - 
P_{W}(t)|dt +
\int_{t_{\times}}^{\infty}
|P_{T}(t)-P_{Power}(t)|dt \nonumber \\ 
& \equiv &
\varepsilon_{W} (t_{\times}) + \varepsilon_{Power} (t_{\times}) 
\label{eq:eps_split}
\end{eqnarray}
%%%%%%%%%%%%%%%%
which is proportional to the gap between the 
true value of the average waiting time 
and the same quantity estimated by the Weibull 
distribution with a power-law tail. 
In the limit of $t_{\times} \to \infty$, 
the difference $\varepsilon$ leads to only the Weibull contribution $\varepsilon_{W}(\infty)$, 
whereas the difference $\varepsilon$ is identical to only the power-law contribution
$\varepsilon_{Power}(0)$ for $t_{\times} \to 0$. 
It is possible to show that there is a specific crossover 
point $t_{\times}$ at which the difference $\varepsilon$ takes 
its minimum. 

Indeed, we first notice that 
the absolute values 
in equation (\ref{eq:eps_split}) 
can be removed by taking into account the relationships between the 
magnitudes of the three distributions 
$P_{T}(t)$, 
$P_{W}(t)$ and $P_{Power}(t)$. 
Then, it is possible to take the derivative of 
$\varepsilon$ with respect to $t_{\times}$ as follows. 
%%%%%%%%%%%%%%
\begin{eqnarray}
\label{epsder}
\frac{d \varepsilon}{dt_{\times}} & = & 
\left\{
\begin{array}{lc}
-P_{W}(t_{\times}) +P_{Power}(t_{\times}) & 
(P_{T} > P_{W}) \land (P_{T} > P_{Power}) \\
%%%%
2P_{T}(t_{\times} ) - 
P_{W}(t_{\times}) +
P_{Power}(t_{\times}) & 
(P_{T} > P_{W}) \land (P_{T} <  P_{Power}) \\
%%%%
-2P_{T}(t_{\times} ) + 
P_{W}(t_{\times}) - 
P_{Power}(t_{\times}) & 
(P_{T} < P_{W}) \land (P_{T} > P_{Power}) \\
%%%%
P_{W}(t_{\times}) - P_{Power}(t_{\times}) & 
(P_{T} < P_{W}) \land (P_{T} <  P_{Power})
\end{array}
\right. 
\end{eqnarray}
%%%%
In order to show that the $\varepsilon$ takes 
its minimum at finite $t_{\times}$, we prove 
that there is a value  of $t_{\times}$ which satisfies 
$d\varepsilon/dt_{\times} = 0$, that is,
%%%%%%%
%%%%%%%%%%%%%%%%%
\begin{eqnarray}
P_{W}(t_{\times}) & = & 
P_{Power}(t_{\times})
\label{eq:cond_nonmonotonic0}
\end{eqnarray}
%%%%%%%%%
for 
$((P_{T} > P_{W}) \land (P_{T} > P_{Power})) 
\lor 
((P_{T} < P_{W}) \land (P_{T} <  P_{Power}))$, 
%%%%
and 
%%%%%%%%%%%
\begin{eqnarray}
2P_{T}(t_{\times}) & = & P_{W}(t_{\times}) + 
P_{Power}(t_{\times})
\label{eq:cond_nonmonotonic}
\end{eqnarray}
%%%%%%%
for 
$((P_{T} > P_{W}) \land (P_{T} <  P_{Power})) \lor 
((P_{T} < P_{W}) \land (P_{T} > P_{Power}))$. 
%%%%%%%%%%%%%%%

Actually, we defined the Weibull 
distribution with a power-law tail to satisfy 
$P_{W}(t_{\times}) = P_{Power}(t_{\times}) = 
P_{T}(t_{\times})$ 
in order to approximate the true empirical distribution. 
In the previous section, we obtained the 
Weibull distribution 
with a power-law tail (\ref{eq:conti_cond}) 
by taking into account the condition 
$P_{W}(t_{\times})=P_{Power}(t_{\times})$, namely, 
the continuity between two curves at 
the crossover point. 
As the value of the empirical distribution 
at $t=t_{\times}$, namely, 
$P_{T}(t_{\times})$ 
is close to 
the theoretical prediction 
$P_{W}(t_{\times})$ (or of course $P_{Power}(t_{\times})$) 
for $m=0.585$, 
the condition 
$P_{W}(t_{\times}) = P_{Power}(t_{\times}) \simeq  
P_{T}(t_{\times})$, 
namely, both 
(\ref{eq:cond_nonmonotonic0}) and 
(\ref{eq:cond_nonmonotonic}) 
are satisfied. 

Therefore, our statement holds true: there exists a crossover 
point $t_{\times}$ at which 
the difference $\varepsilon$ takes its minimum. 
The non-monotonicity of the curve of the average 
waiting time is nothing but an effect of the fact that 
the difference $\varepsilon$ is minimized for the 
intermediate value of $t_{\times}$. 

This is another 
intuitive explanation 
for the non-monotonic behavior of 
the average waiting time as a 
function of $t_{\times}$. 
However, to determine the 
value $t_{\times}$, 
we need more information. 
Then, 
we use the fact 
discussed in the previous subsection, 
namely, 
the difference 
between 
the empirical distribution 
and the Weibull distribution with 
a power-law tail 
is proportional to 
the difference of the 
two distinct areas 
$\mathcal{A}_{1}-\mathcal{A}_{2}$. 
The difference $\epsilon$ 
is written 
in terms 
of the distribution 
$P_{W}(t)$ and 
$P_{Power}(t)$ as follows. 
%%%%%%%%
\begin{eqnarray}
\label{epsilon}
\epsilon & = & 
\int_{0}^{\infty}
\left\{
P_{W}(t) - 
P_{Power}(t) 
\right\}dt \nonumber \\
\mbox{} & = &  
\frac{m}{a} 
\int_{0}^{\infty}
t^{m-1} 
{\rm e}^{-\frac{t^{m}}{a}} dt
- 
\frac{m}{a} 
t_{\times}^{m +\gamma -1}
{\rm e}^{-\frac{t_{\times}^{m}}{a}}
\int_{0}^{\infty}
t^{-\gamma} dt 
\end{eqnarray}
%%%%%%%
where we used the 
explicit forms of 
the distributions 
$P_{W}(t)$ and $P_{Power}(t)$ 
to obtain the second line 
of the above equation. 
%%%%%%
Then, we take the derivative 
of 
$\epsilon$ 
with respect to 
$t_{\times}$ and set it to zero, 
that is 
$\partial \epsilon/\partial t_{\times} =0$,  
in order to obtain the necessary condition to let $\epsilon$ take 
its maximum at $t_{\times}$. 
Then, we have
%%%%%%%
\begin{eqnarray}
\label{tcross}
t_{\times}^{*} & = & 
\left\{
\frac{a}{m}
(m + \gamma -1) 
\right\}^{\frac{1}{m}}
\label{eq:formulaTx}. 
\end{eqnarray}
%%%%%%%%%
This 
value 
$t_{\times}^{*}$ might be a candidate 
to give an optimal crossover point for 
which 
the gap of the average waiting time 
$w$ for the empirical distribution and 
for the Weibull distribution with a 
power-law tail 
is minimized. 
To compare the value for 
the true parameter set 
$(m,a,\gamma)$ obtained 
from the empirical data analysis with 
that obtained in the previous subsection 
$C\simeq24000$ [{\rm s}], 
we substitute the values 
$m=0.585, 
a=49.63$ and $\gamma = 4.67$ into 
the above expression 
(\ref{eq:formulaTx}) and 
immediately obtain 
%%%%%
\begin{eqnarray}
\label{tcross1}
t_{\times}^{*} & \simeq  & 
23538.3 \,[{\rm s}].
\end{eqnarray}
%%%%%
This result is very close to 
the value $C\simeq24000$ in 
the previous section. 
Inserting the above 
crossover point $t_{\times}^{*}$ with 
the other parameters 
$(m,a,\gamma)$ estimated by 
empirical data analysis 
into 
the expression (\ref{eq:formulaW}), 
we obtain the average waiting time 
for the Sony Bank rate as 
$w = 46.25$ [min]. 
Then, the gap $\Delta$ is estimated as 
$\Delta w =49.19 - 46.25 = 2.94$ [min]. 
Therefore, 
the correction obtained by 
modifying 
the crossover point 
reduces the gap $\Delta w$ 
between the empirical and the theoretical 
predictions 
from $\Delta w = 3.53$ [min] to 
$\Delta w = 2.94$ [min]

Thus, we obtained a formula to 
determine the 
appropriate (and may be 
an optimal) crossover point 
$t_{\times}^{*}$ 
for our proposed 
first-passage time 
distribution, 
that is, 
the Weibull distribution 
with a power-law tail. 
It is 
important to 
stress that 
formula (\ref{eq:formulaTx}) is rather
general and can be always applied to data
described by a Weibull distribution with
power-law tail.
%%%%%%%%%%%%%%%%%%%%%%%%%%%%%%%%%%%%%%%%%%%%%%%%%%%%%%%%%%%%%%%%
\subsection{On the sign of the second derivative 
of $\varepsilon$ to confirm that $\varepsilon$ 
takes its minimum at $t_{\times}$}
%%%%%%%%%%%%%%%%%%%%%%%%%%%%%%%%%%%%%%%%%%%%%%%%%%%%%%%%%%%%%%%
In order to confirm that $\varepsilon$ takes its minimum (not its maximum) 
at $t_{\times}$, we can evaluate the sign of the second derivative 
of $\varepsilon$ with respect to $t_{\times}$, that is, $d^{2} \varepsilon/dt_{\times}^{2}$. 
One can label the cases in equation (\ref{epsder}) as follows: 
%%%%%%%%%%%%%%
\begin{eqnarray*}
\frac{d \varepsilon}{dt_{\times}} & = & 
\left\{
\begin{array}{lc}
-P_{W}(t_{\times}) +P_{Power}(t_{\times}) & 
(P_{T} > P_{W}) \land (P_{T} > P_{Power}) \,\,\,
{\bf (A)} \\
%%%%
2P_{T}(t_{\times} ) - 
P_{W}(t_{\times}) +
P_{Power}(t_{\times}) & 
(P_{T} > P_{W}) \land (P_{T} <  P_{Power})\,\,\,
{\bf (B)} \\
%%%%
-2P_{T}(t_{\times} ) + 
P_{W}(t_{\times}) - 
P_{Power}(t_{\times}) & 
(P_{T} < P_{W}) \land (P_{T} > P_{Power})\,\,\,
{\bf (C)} \\
%%%%
P_{W}(t_{\times}) - P_{Power}(t_{\times}) & 
(P_{T} < P_{W}) \land (P_{T} <  P_{Power})\,\,\,
{\bf (D)}
\end{array}
\right. 
\end{eqnarray*}
%%%%%%%%%%%%%%%
It should be kept in mind that the empirical data should fall 
in one of the above four categories: {\bf case A}, {\bf case B}, 
{\bf case C}, {\bf case D}. 
In addition, we should notice that 
each conjuncted condition in {\bf case A} is opposite to 
{\bf case D}, and each conjuncted condition in
{\bf case B} is opposite to 
{\bf case C} with respect to the sign of the second 
derivative of $\varepsilon$ at $t_{\times}$. 
Therefore, we should check whether the second 
derivative of $\varepsilon$ takes positive 
value or not in each case. 
%Then, two of the above four cases might be rejected 
%by simple evaluation of the sign of the second derivative. 

\begin{figure}[ht]
\begin{center}
\includegraphics[width=11cm]{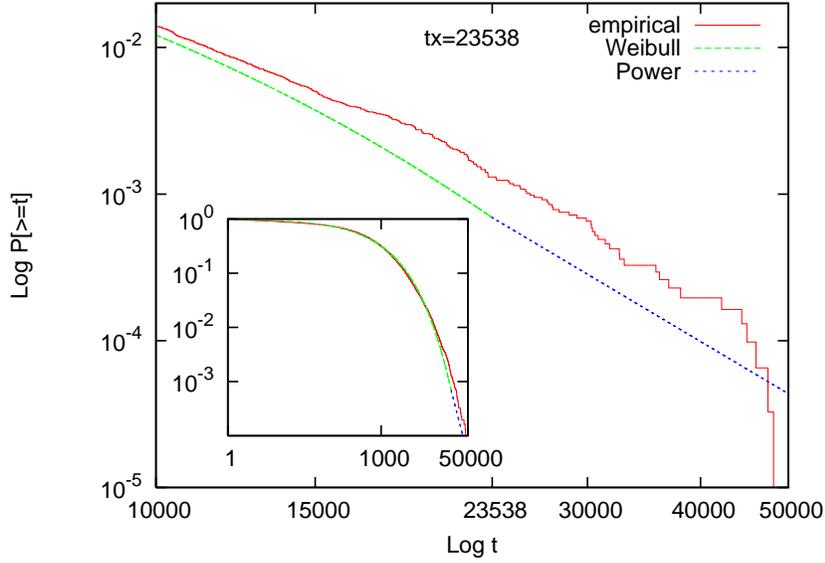}
\end{center}
\caption{\footnotesize 
The empirical data analysis 
suggests 
$(P_{T} > P_{W}) \land 
(P_{T} > P_{power})$ is true ({\bf case A}). 
The inset is a full complementary 
cumulative distribution function.}
\label{fig:fginset}
\end{figure}

For any case, we need the derivatives 
$dP_{W}/dt_{\times}, 
dP_{power}/dt_{\times}$ and 
$dP_{T}/dt_{\times}$. 
The last one $dP_{T}/dt_{\times}$ denotes a derivative 
of the empirical distribution at $t_{\times}$ and we 
should evaluate of the derivative numerically 
from the emprical data. 
However, 
the empirical data analysis 
suggests 
$(P_{T} > P_{W}) \land 
(P_{T} > P_{power})$ is true ({\bf case A})
(see Figure \ref{fig:fginset}). 
Actually, 
we do not need the evaluation of $dP_{T}/dt_{\times}$ and 
we need only the above first two derivatives. 
We obtain them analytically 
%%%%%%
\begin{eqnarray}
\frac{dP_{W}}{dt_{\times}} \equiv 
\frac{dP_{W}}{dt}
{\Biggr |}_{t=t_{\times}} & = & 
\frac{m}{a}t_{\times}^{m-2}
{\rm e}^{-\frac{t_{\times}^{m}}{a}}
\left(
m-1-
\frac{m}{a} t_{\times}^{m}
\right) \\
%%%%%
\frac{dP_{power}}{dt_{\times}} \equiv 
\frac{dP_{power}}{dt}
{\Biggr |}_{t=t_{\times}} & = &  
-\frac{\gamma m}{a} 
t_{\times}^{m-2}
{\rm e}^{-\frac{t_{\times}^{m}}{a}} < 0.
\end{eqnarray}
%%%%%
%%%%%%%%%%%%%%%%%%%%%%%
%%%%%%%%%%%%%%%%%%%%%%%%%
We should notice that the sign of $dP_{power}/dt_{\times}$ 
is negative for any choice of the parameters $\gamma,m,a$ 
and $t_{\times}$. 
However, the sign of the $dP_{W}/dt_{\times}$ 
depends on the parameters. 
For instance, for $t_{\times}=0$, 
$dP_{W}/dt_{\times} >0$ for $m>1$, whereas,  
$dP_{W}/dt_{\times} < 0$ for $m < 1$. 

Then, we evaluate the differences: 
%%%%%%
\begin{eqnarray}
\frac{dP_{power}}{dt_{\times}} - 
\frac{dP_{W}}{dt_{\times}} & = &   
\frac{m}{a} 
t_{\times}^{m-2} 
{\rm e}^{-\frac{t_{\times}}{a}} 
D(t_{\times})
\end{eqnarray}
%%%%
where we defined 
\begin{eqnarray}
D(t_{\times}) & \equiv & 
\frac{m}{a} 
t_{\times}^{m} -\gamma -m +1. 
\end{eqnarray}
%%%%%%%%%%%%%%%%%%%%
\begin{figure}[ht]
\begin{center}
\includegraphics[width=10cm]{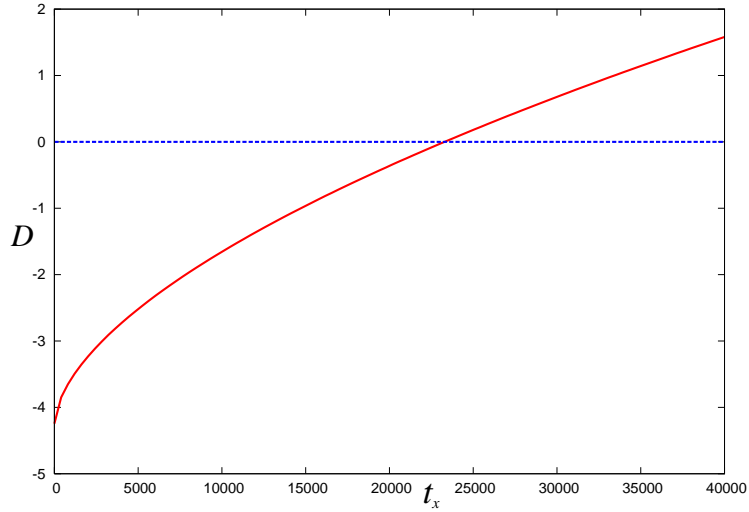}
\end{center}
\caption{\footnotesize 
The behavior of the function $D$ as a function of $t_{\times}$. 
We set 
$m=0.585, a=49.63$ and 
$\gamma=4.67$ as 
the Sony Bank data.}
\label{fig:fg1new}
\end{figure}
%%%%%%%%%%%%%%%%%%%%%%
In Figure \ref{fig:fg1new}, we plot the $D$ as 
a function of 
$t_{\times}$ for 
parameter values
$m=0.585, a=49.63$ and 
$\gamma=4.67$ as in the case of 
the Sony Bank data. 
From this figure, we 
find that 
at some critical point $t_{\times}^{*} \simeq 23538.3$ [{\rm s}], 
the sign of 
the function $D$ changes. 
%%%%%%%%%%%%%%%%%%%%%%%%%
By taking into account 
the fact that 
the crossover point 
used here is 
$C \equiv t_{\times}=
24000 > t_{\times}^{*}$, 
we conclude that 
%%%%%%%%%%%%%
\begin{eqnarray}
\frac{dP_{power}}{dt_{\times}} >  
\frac{dP_{W}}{dt_{\times}}. 
\end{eqnarray}
%%%%%
Therefore, considering that 
the empirical data analysis 
suggests 
$(P_{T} > P_{W}) \land 
(P_{T} > P_{power})$ is true ({\bf case A}), 
we prove that 
$\varepsilon$ takes 
its minimum 
at $t_{\times}$. 
%%%%%%%%%%%%%%%%%%%%%%%%%%

In order to discuss $t_{\times}$ in more detail, 
we start from equation (\ref{epsilon})
and take the derivative of $\epsilon$
with respect to $t_{\times}$
%%%%%%%%%%%%%%%%%%%%%%%%%%%%%%%%%%%%%%
\begin{eqnarray}
\frac{\partial \epsilon}{\partial t_{\times}}
 & = & -\frac{m}{a} 
\,{\rm e}^{-\frac{t_{\times}^{m}}{a}}
t_{\times}^{m +\gamma -2}
\left\{
(m +\gamma -1)-\frac{m}{a}t_{\times}^{m}
\right\} \int_{0}^{\infty}
t^{-\gamma} dt 
\end{eqnarray}
%%%%%%%%%%%%%%%%%%%%%%%%%%%%%%%%%%%%%%%%%
By taking $\partial \epsilon/\partial t_{\times}=0$,
one obtains (\ref{tcross}) as the solution $t_{\times}$ 
and the value for the empirical data is given by (\ref{tcross1}). 
To confirm that the solution $t_{\times}$ gives the maximum of 
the $\epsilon$, 
we check the sign of the second derivative of
$\epsilon$. 
We find 
%%%%%%%%%%%%%%%%%%%%%%%%%%%%%%%%%%%%%%%%
\begin{eqnarray}
\mbox{} & \mbox{} & \frac{\partial^2 \epsilon}{\partial t_{\times}^2}
 = 
-\frac{m}{a} 
{\rm e}^{-\frac{t_{\times}^{m}}{a}}
t_{\times}^{m +\gamma -3}  \nonumber \\
\mbox{} & \times & 
\left\{
\left( \frac{m}{a} \right)^{2}t_{\times}^{2m}
-\frac{mt_{\times}}{a}
\left(
m+1\right)
+(m +\gamma -1)(m +\gamma -2)
\right\} \int_{0}^{\infty}
t^{-\gamma}dt. 
\end{eqnarray}
%%%%%%%%%%%%%%%%%%%%%%%%%%%%%%%%%%%%%%%%%%%%%
Therefore, by replacing 
(\ref{tcross}), namely, $t_{\times}=t_{\times}^{*}$ 
into the above expression, we get
%%%%%%%%%%%%%%%%%%%%%%%%%%%%%%%%%%%%%%%%%%%%%%
\begin{eqnarray}
\frac{\partial^2 \epsilon}{\partial t_{\times}^2}
{\Biggr |}_{t_{\times}=t_{\times}^{*}} 
& = & 
m(m +\gamma -1)^{2}
{\rm e}^{-\frac{m +\gamma -1}{m}} 
\left\{
\frac{a}{m}
(m + \gamma -1) 
\right\}^{\frac{\gamma -3}{m}} \int_{0}^{\infty}
t^{-\gamma}dt
\end{eqnarray}
%%%%%%%%%%%%%%%%%%%%
and we conclude that 
the $\epsilon$ takes 
its maximum at $t_{\times}=t_{\times}^{*}$, 
that is,  
$\partial^2 \epsilon/\partial t_{\times}^2> 0$ 
for the solution of $\partial \epsilon/\partial t_{\times}=0$. 
%%%%%%%%%%%%%%%%%%%%%%%%%%%%%%%%%%%%%%%%%%%%%
Therefore $\epsilon$ has a minimum at $t^*_{\times}$.
%%%%%%%%%%%%%%%%%%%%%%%%%%%%%%%%%%%%%%%%%%%%%%%%%%%%%%%%%%%%%%
\section{Application to BTP future data}
%%%%%%%%%%%%%%%%%%%%%%%%%%%%%%%%%%%%%%%%%%%%%%%%%%%%%%%%%%%%%%%
It is now 
interesting to see what happens when we apply the Weibull 
distribution with a power-law tail to 
another financial data set and a different random variable. 
As mentioned several times, in the BTP future case, we study
intertrade durations and not first-passage times. 
In this section, we 
evaluate the average waiting time 
for the BTP future. 
Then, we investigate 
to what extent  
our formulation is applicable and 
we also discuss the limits of 
that formulation. 
%%%%%%%%%%%%%%%%%%%%%%%%%%%%%%%%%%%%%%%%%%%%%%%%%%%%%%%%%%%%%
\subsection{Weibull-paper analysis for 
the BTP future}
%%%%%%%%%%%%%%%%%%%%%%%%%%%%%%%%%%%%%%%%%%%%%%%%%%%%%%%%%%%%%
To evaluate the average waiting time $w$ 
for 
the Weibull distribution with 
a power law-tail, 
we estimate the parameters 
$a,m$ and $t_{\times}$ from 
the available empirical data. 
To this purpose, we 
carry out the so-called Weibull-paper analysis.
%%%%%%%%%%%%%%%%%%%%%%%
\begin{figure}[ht]
\begin{center}
\includegraphics[width=11cm]{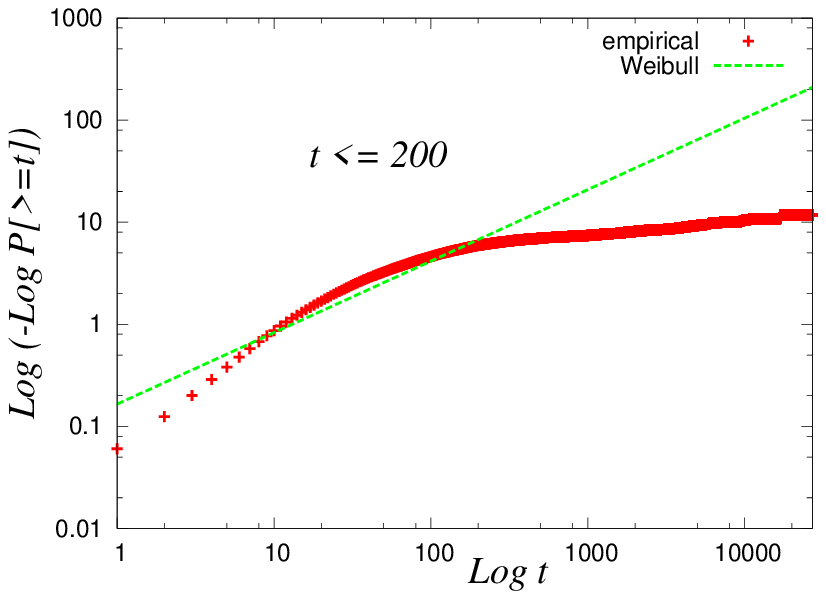}
\end{center}
\caption{\footnotesize 
Weibull paper analysis for the data $t\le200\,[{\rm s}]$.}
\label{fig:fgD}
\end{figure} 
%%%%%%%%%%%%%%%%%%%%%%%%%
We show the result in 
Fig. \ref{fig:fgD}.
To produce this Weibull 
paper, 
we used data up to $t=200\,[{\rm s}]$ 
(about $99.7 \,\%$ of the whole data set). 
From this figure, 
we find that 
there are apparent gaps between 
the empirical plot and 
the Weibull paper (straight line). 
Nevertheless, 
from the Weibull paper analysis, 
we obtain 
$m=0.85$ and 
$a=10.02$. 
For these parameters, 
the average waiting time 
estimated 
by the renewal-reward theorem 
leads to 
%%%%%%
\begin{eqnarray}
w & = & 
a^{\frac{1}{m}}
\frac{\Gamma (\frac{2}{m})}
{\Gamma (\frac{1}{m})}  \simeq 
30.0 \,[{\rm s}]
\label{eq:theoryBTP}
\end{eqnarray}
%%%%%%%%%%%%
whereas, 
by sampling from the 
empirical data, 
we obtain $\langle t \rangle  \simeq 16.5511\,[{\rm s}]$ and 
%%%%
\begin{eqnarray}
w & = & 
\frac{\langle t^{2} \rangle}
{2 \langle t \rangle} \simeq 530.1 \,[{\rm s}] = 
8.8\, [{\rm min}].
\end{eqnarray}
%%%%%%%%%%%
since from the empirical data one finds $w > \langle t \rangle$,
the inspection paradox occurs for the BTP future data.
Moreover, the empirical result is far from 
the theoretical 
prediction (\ref{eq:theoryBTP}).
The reason for the large gap might come from 
the bad fit of 
the empirical data 
by means of a pure Weibull distribution. 

\begin{figure}[ht]
\begin{center}
\includegraphics[width=11cm]{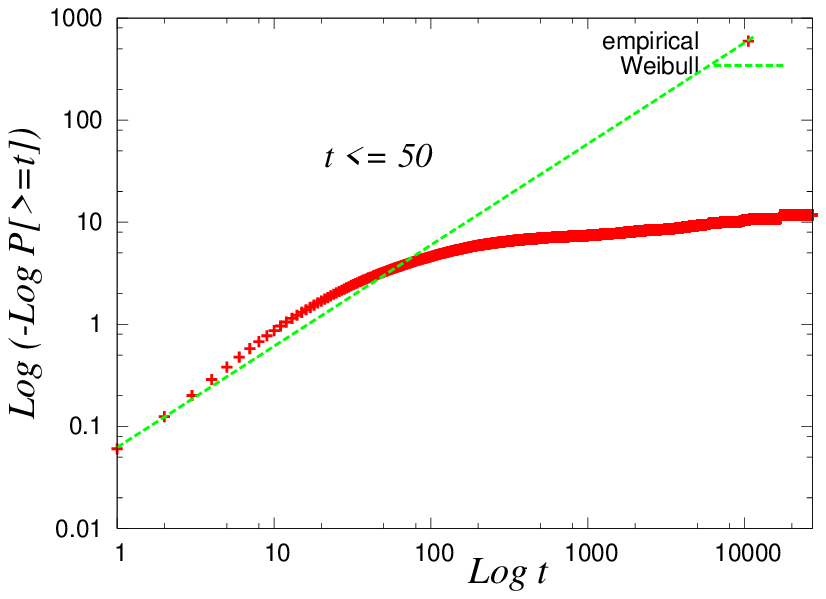}
\end{center}
\caption{\footnotesize 
Weibull paper analysis for the data $t\le50\,[{\rm s}]$.}
\label{fig:fgE}
\end{figure}

We next reduce the range to fit the data by Weibull paper 
analysis from $t=200\,[{\rm s}]$ to $t=50\,[{\rm s}]$ which is 
about $96.1\, \%$ of whole data points. 
In Fig. \ref{fig:fgE}, 
we display the Weibull paper and obtain 
the parameters as 
$m=0.99$ and $a=16.49$.  By making use of 
these parameters, 
the theoretical prediction of 
the average waiting time leads to $w \simeq 16.70\,[{\rm s}]$. 
This value is very close to  
the value of 
the first moment for the empirical data $\langle t \rangle$. 
This result tells us that 
the Weibull paper analysis for the data point up to 
$t=50$ gives almost the same prediction as an exponential distribution 
for the duration of the BTP future. 
%%%%%%%%%%%%%%%%%%%%%%%
%%%%%%%%%%%%%%%%%%%%%%%%%
%%%%%%%%%%%%%%%%%%%%%%%%%%%%%%%%%%%%%%%%%%%%%%%%%%%%%%%%%%%%%
\subsection{Weibull distribution with 
a power-law tail for the BTP futures}
%%%%%%%%%%%%%%%%%%%%%%%%%%%%%%%%%%%%%%%%%%%%%%%%%%%%%%%%%%%%%
We now evaluate the 
optimal crossover point 
$t_{\times}^{*}$ from 
the parameters obtained by empirical data analysis. 
Inserting 
these values 
$m=0.70, 
a=6.05$ and $\gamma=1+\beta =1.96$ for 
$t=200$ data points into 
our formula (\ref{eq:formulaTx}), 
we have 
%%%%%%%%%%
\begin{eqnarray}
t_{\times}^{*} & \simeq & 
44.9 \,[{\rm s}].
\end{eqnarray}
%%%%
On the other hand, 
when we use the $t=50$ data points, 
we use the values
$m=0.99, a=16.49$ and $\gamma=1.96$ into 
the expression (\ref{eq:formulaTx}) and obtain
%%%%%%%%%%
\begin{eqnarray}
t_{\times}^{*} & \simeq & 33.5 \,[{\rm s}].
\end{eqnarray}
%%%%%%%%%%
We next evaluate the average waiting time 
$w$ by using the 
formula (\ref{eq:formulaW}) which was corrected by means of 
the power-law tail effect. 
%%%%%%%%%%%
%%%%%%%%%%%%%%%%%%%%%%%
\begin{figure}[ht]
\begin{center}
\includegraphics[width=11cm]{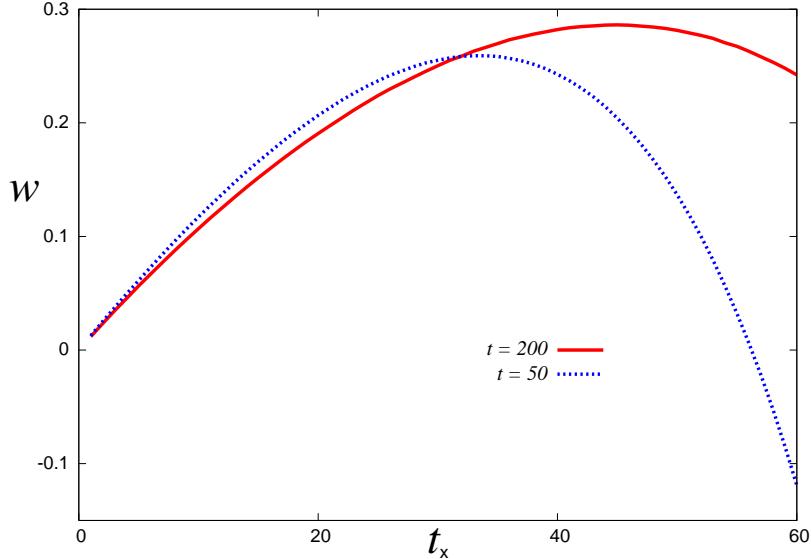}
\end{center}
\caption{\footnotesize 
The corrected average waiting time $w$ 
for the BTP future as a function of $t_{\times}$. 
We set $\gamma = 1.96$. 
For two cases of the choice for 
the number of 
data points $t$, namely, for $t=200$
($m=0.70, a=6.05$) and $50$
($m=0.99, a=16.49$),  the $w$ are plotted.}
\label{fig:fg7}
\end{figure} 
%%%%%%%%%%%%%%%%%%%%%%%
In Fig. \ref{fig:fg7}, we 
plot the corrected average waiting time $w$ 
for the BTP future as a function of $t_{\times}$ 
for both cases of 
$t$, namely, for $t=200$
($m=0.70, a=6.05$) and $50$ 
($m=0.99, a=16.49$). 
From this figure, we see that, 
at the predicted 
optimal crossover point $t_{\times}^{*}$, 
both curves take their maximum, 
however, the values of the average waiting 
time are lower than (to make matters worse, for large 
$t_{\times}$ it becomes 
negative) those estimated by a pure Weibull distribution. 
This is because the second terms appearing in 
both numerator and denominator in the formula becomes negative 
for the parameter range of $\gamma<2$. 
Thus, we conclude that the BTP future 
has too heavy a tail ($\gamma=1.96$) 
to correct the average waiting time 
by using the 
formula (\ref{eq:formulaW}). 
This is a limitation of our formula for 
the average waiting time for financial data. 
%%%%%%%%%%%%%%%%%%%%%%%%%%%%%%%%%%%%%%%%%%%%%%%%%%%%%%%%%%%%%%
\section{Summary and discussion}
%%%%%%%%%%%%%%%%%%%%%%%%%%%%%%%%%%%%%%%%%%%%%%%%%%%%%%%%%%%%%%%%
In this paper, we have compared a Weibull distribution 
and a Mittag-Leffler distribution. Then, 
two relevant statistics, 
namely, 
the average waiting time and 
the Gini index have been studied in both cases. 
Our theoretical analysis revealed that the average waiting 
time diverges linearly as a function of the cut-off parameter $t_{\rm max}$
for the Mittag-Leffler distribution. 
This fact implies a more difficult treatment 
to check the validity of modeling the market renewal process
by means of the Mittag-Leffler distribution. 
On the other side, 
the Gini index for 
the Mittag-Leffler survival 
function is free from this kind of 
divergence because 
the tail part of 
the duration distribution 
does not contribute to the value so much.  
We also find that a Weibull distribution 
with a power-law tail is an efficient 
way to describe renewal processes in markets 
with a long duration such as the Sony Bank USD/JPY 
exchange rate seen as a first-passage process. We conclude 
that the Weibull distribution with a power-law tail is more suitable 
to evaluate the relevant statistics for financial 
markets with a long duration. 
By considering the intuitive explanation of 
the non-monotonic behavior of 
the corrected average waiting time 
as a function of 
the crossover point, 
we obtained 
a useful formula 
to decide the appropriate (and might be an optimal) 
crossover point $t_{\times}^{*}$. 
In fact, 
we could reduce the gap 
of the average waiting time 
$\Delta w$ between 
the theoretical and empirical 
data analysis 
from $\Delta w = 4.57$ [min]
(for a pure Weibull distribution) 
to $\Delta w = 2.94$ [min] 
by evaluating the average waiting time 
with the optimal crossover point 
$t_{\times}^{*}$ and 
the parameter set 
$(m,a,\gamma)$ 
obtained by the empirical data analysis of 
the Sony Bank rate. 
To investigate the limitation of 
our distribution 
to describe the other financial data, 
we applied our distribution to 
the BTP future. 
We found from the Weibull paper analysis that 
for the short range duration regime, 
there exist apparently gaps between 
the empirical and our proposed distributions. 
To make matters worse, we concluded that 
the BTP future has too heavy tail to 
obtain the correction for the average waiting time 
by means of our formula (\ref{eq:formulaW}). 
From these observations, we could say 
that our proposed distribution, namely, 
the Weibull distribution with 
a power-law tail is applicable to 
the financial data having the following 
two properties. 
%%%%%%%%%%%
\begin{itemize}
\item
In short duration regime, 
it follows a Weibull-law. 
%%%%
\item
It does not have too heavy tail, namely, $\gamma > 3$ should be needed. 
\end{itemize}
%%%%%%%%%
If the above two conditions hold in the financial data, 
the duration of the data 
might be well described by our proposed 
distribution.  

We hope that our proposed method will be widely used as a 
powerful candidate to describe the duration 
in financial data having the above two properties. 
%%%%%%%%%%%%%%%%%%%%%%%%%%%%%%%%%%%%%%%%%%%%%%%%%%%%%%%%%%%%%%%%%
%%                       Acknowledgements                       %%
%%%%%%%%%%%%%%%%%%%%%%%%%%%%%%%%%%%%%%%%%%%%%%%%%%%%%%%%%%%%%%%%%%
\section*{Acknowledgment}
E.S. is grateful to JSPS for a short-term fellowship in Japan
at the International Christian University, Tokyo, in the
group of Prof. T. Kaizoji during which
this paper has been discussed. 
J.I. was 
financially supported 
by {\it Grant-in-Aid 
Scientific Research on Priority Areas 
``Deepening and Expansion of Statistical Mechanical Informatics (DEX-SMI)" 
of The Ministry of Education, Culture, 
Sports, Science and Technology (MEXT)} 
No. 18079001. 
N.S. would like to acknowledge useful discussion 
with Shigeru Ishi, President of the Sony bank.
The authors wish to thank Prof. T. Kaizoji
for useful discussion.
%%%%%%%%%%%%%%%%%%%%%%%%%%%%%%%%%%%%%%%%%%%%%%%%%%%%%%%%%%%%%%%%%
%%                       References                            %%
%%%%%%%%%%%%%%%%%%%%%%%%%%%%%%%%%%%%%%%%%%%%%%%%%%%%%%%%%%%%%%%%%%

\end{document}